# Analytical model for nonlinear response of carbon nanotubes enhanced by a plasmonic metamaterial

A. Chipouline[1], S. Sugavanam[1], V. A. Fedotov[2], A. E. Nikolaenko[2] and T. Pertsch[1],
[1]*Institute of Applied Physics, Friedrich Schiller University Jena, Max-Wien-Platz 1, D-07743 Jena, Germany*
[2]*Optoelectronics Research Centre and Centre for Photonic Metamaterials, University of Southampton, SO17 1BJ, UK*

**Abstract**

We present an analytical model describing complex dynamics of a hybrid nonlinear system consisting of interacting carbon nanotubes (CNT) and a plasmonic metamaterial.

Our model is based on the set of coupled equations, which incorporates well-established density matrix formalism appropriate for quantum systems (CNT are described as a two level system) and harmonic-oscillator approach ideal for modelling sub-wavelength plasmonic and optical resonators.

We show that the saturation nonlinearity of CNT increases multifold in the resonantly enhanced near field of a metamaterial. In the framework of our model, we discuss the effect of inhomogeneity of the CNT layer (band gap value distribution) on the nonlinearity enhancement. It is shown, that the Purcell effect is indistinguishable from the field enhancement and is described by the same phenomenological constant.

## I. Introduction

With the development of nanotechnology it has become possible to engineer and study novel hybrid quantum-classical nanoscale systems with enhanced nonlinear response such as, for example, a plasmonic metamaterials combined with carbon nanotubes (CNT) [1]. In such a system the metamaterial structure works as a light concentrator enhancing local optical fields, which are coherent with the incident field and also affect the response dynamics of CNT. In the resonance case the intensity of the local fields can become significantly higher then that of the incident wave and therefore substantially affect the dynamics of CNT response. Here, we investigate this effect theoretically using our recently developed general model that describes hybridized eigen modes of a coupled classical/quantum system [2].

## II. Model

In the frame of our model the near-filed interaction between the classical and quantum components of the hybrid system is accounted for using the dipole-dipole approximation [2], while plasmonic nano-resonators of the metamaterial are described by a set of harmonic oscillator equations [3]. The model is base on the following master set of equations:

$$\begin{cases} \dfrac{d\rho_{12}}{dt} + \rho_{12}\left(\dfrac{1}{\tau_2} + i(\omega - \omega_{21})\right) = \dfrac{i\alpha_x x^* N}{\hbar} + \dfrac{i\mu_{CNT} A^* N}{\hbar} \\ \dfrac{dN}{dt} + \dfrac{(N - N_0)}{\tau_1} = \dfrac{i\alpha_x (x\rho_{12} - x^*\rho_{12}^*) + i\mu_{CNT}(A\rho_{12} - A^*\rho_{12}^*)}{2\hbar} \\ 2(\gamma - i\omega)\dfrac{dx}{dt} + (\omega_0^2 - \omega^2 - 2i\omega\gamma)x = \alpha_\rho \rho_{12}^* + \chi A \end{cases} \quad (1)$$

Here $A$ is the amplitude of the electric field, $N = \rho_{22} - \rho_{11}$, $\rho_{22}$, $\rho_{11}$ and $\rho_{12}$, $\rho_{12}^*$ are the diagonal and slow amplitude non-diagonal matrix density elements, respectively; $x$ is the amplitude of electronic oscillations in the metamaterial, $\chi^{-1}$ is the effective kinetic inductance of electrons; $N_0 = \dfrac{W\tau_1 - 1}{W\tau_1 + 1}$ is the population inversion due to pump (in the absence of pump $N_0 = -1$), $N_0 > 0$ corresponds to the regime of amplification, $N_0 < 0$ - to losses; $\tau_2$ and $\tau_1$ are the constants describing phase and energy relaxation processes due to the interaction with a thermostat; both eigen frequencies $\omega_{21}$ and $\omega_0$ are the resonance frequencies of CNT and metamaterial respectively and can vary independently; $\alpha_x = \dfrac{\mu_{CNT}\mu_{MM}}{R^3}$, $\alpha_\rho = \dfrac{\mu_{CNT}\chi}{R^3}$, $\mu_{CNT}$ is the CNT dipole moment for the band gap transition, $\mu_{MM}$ is the effective metamaterial dipole moment, $R$ is the effective distance between CNT and metamaterial structure.

### III. Applications of model for nonlinear response enhancement

The nonlinearity of CNT appears due to the saturation induced by the direct pumping and basically requires neither positive $N_0$ nor nano-resonator. The enhancement of the nonlinearity is caused by the addition pumping channel, where - external field transfers energy to CNT through the nano-resonator. This is described by the term $\dfrac{i\alpha_x x^* N}{\hbar}$ in the first equation of system (1).

In order to solve (1) analytically we assume that the plasmonic resonator is driven mainly by the external electric field:

$$\begin{cases} \dfrac{d\rho_{12}}{dt} + \rho_{12}\left(\dfrac{1}{\tau_2} + i(\omega - \omega_{21})\right) = \dfrac{i\alpha_x x^* N}{\hbar} + \dfrac{i\mu_{CNT} A^* N}{\hbar} \\ \dfrac{dN}{dt} + \dfrac{(N - N_0)}{\tau_1} = \dfrac{i\alpha_x (x\rho_{12} - x^*\rho_{12}^*) + i\mu_{CNT}(A\rho_{12} - A^*\rho_{12}^*)}{2\hbar} \\ 2(\gamma - i\omega)\dfrac{dx}{dt} + (\omega_0^2 - \omega^2 - 2i\omega\gamma)x = \chi A \end{cases} \quad (2)$$

and then consider the case of CW excitation:

$$\begin{cases} \rho_{12}\left(\dfrac{1}{\tau_2}+i(\omega-\omega_{21})\right)=\dfrac{i\alpha_x x^* N}{\hbar}+\dfrac{i\mu_{CNT}A^* N}{\hbar} \\ \dfrac{(N-N_0)}{\tau_1}=\dfrac{i\alpha_x(x\rho_{12}-x^*\rho_{12}^*)+i\mu_{CNT}(A\rho_{12}-A^*\rho_{12}^*)}{2\hbar} \\ (\omega_0^2-\omega^2-2i\omega\gamma)x=\chi A \end{cases} \quad (3)$$

Our goal is to express $\rho_{12}^*$ and find the effective dielectric constant of CNT layer $\varepsilon_{CNT}=\varepsilon_1+i\varepsilon_1=1+4\pi\left(n_{CNT}\dfrac{\mu_{CNT}\rho_{12}^*}{A}+n_{MM}\dfrac{qx}{A}\right)$ as a function of all parameters and intensity (the second term in brackets appears due to the metamaterial itself); here $n_{CNT}$ and $n_{MM}$ are the concentrations of CNT and metamolecules respectively. The imaginary part of the dielectric constant is responsible for the losses and its intensity and frequency dependence could be compared with the experiments.

In order to make expression for the dielectric constant more compact we have introduced several new notations:

$$\begin{cases} R_x=\omega_0^2-\omega^2-2i\omega\gamma; \quad |R_x|^2=(\omega_0^2-\omega^2)^2+4\omega^2\gamma^2 \\ R_\rho=1+i(\omega-\omega_{21})\tau_2; \quad |R_\rho|^2=1+(\omega-\omega_{21})^2\tau_2^2 \\ \sigma=\dfrac{\alpha_x\chi}{\mu_{CNT}} \\ F_1=\omega_0^2-\omega^2-2\omega\gamma(\omega-\omega_{21})\tau_2 \\ F_2=\omega_0^2-\omega^2+2\omega\gamma(\omega-\omega_{21})\tau_2 \\ F_3=\omega_0^2-\omega^2 \\ A\rho_{12}-A^*\rho_{12}^*=\dfrac{2iN\tau_2\mu|A|^2}{\hbar|R_\rho|^2}\left(1+\dfrac{\sigma F_1}{|R_x|^2}\right) \\ x\rho_{12}-x^*\rho_{12}^*=\dfrac{i2\chi N\tau_2\mu|A|^2}{\hbar|R_x|^2|R_\rho|^2}(\sigma+F_2) \\ |A_s|^2=\dfrac{|A_{s,0}|^2|R_\rho|^2}{1+\dfrac{\sigma(\sigma+F_3)}{|R_x|^2}}; \quad |A_{s,0}|^2=\dfrac{\hbar^2}{\mu\tau_1\tau_2} \\ N=\dfrac{N_0}{1+\dfrac{|A|^2}{|A_s|^2}}=\dfrac{N_0}{1+S}; \quad S=\dfrac{|A|^2}{|A_s|^2}=S_0\left[\dfrac{1+\dfrac{\sigma(\sigma+F_3)}{|R_x|^2}}{|R_\rho|^2}\right] \\ S_0=\dfrac{|A|^2}{|A_{s,0}|^2}; \quad \rho_{12}=\dfrac{-iN\tau_2\mu R_\rho}{\hbar|R_\rho|^2}\left(1+\dfrac{\sigma}{R_x^*}\right)A \end{cases} \quad (4)$$

Here $S_0$ and $S$ account for the saturation and enhanced saturation respectively. Taking into account (4) we obtain the following final expression for the dielectric constant:

$$\begin{cases} \varepsilon_{CNT} + \varepsilon_x = \varepsilon_1 + i\varepsilon_2 \\ \varepsilon_1 = \varepsilon_{1,0} + \dfrac{4\pi n_{CNT}\tau_2\mu^2 N_0}{\hbar |R_\rho|^2 (1+S)}\left((\omega - \omega_{21})\tau_2 + \dfrac{2\omega\gamma\sigma}{|R_x|^2}\right) + \dfrac{4\pi n_{MM} q\chi(\omega_0^2 - \omega^2)}{|R_x|^2} \\ \varepsilon_2 = -\dfrac{4\pi n_{CNT}\tau_2\mu^2 N_0}{\hbar |R_\rho|^2 (1+S)}\left(1 + \dfrac{\sigma F_1}{|R_x|^2}\right) + \dfrac{8\pi q n_{MM}\chi\omega\gamma}{|R_x|^2} \end{cases} \quad (5)$$

It represents the solution of the problem in framework of our model and approximations made. Let us assess the validity of the model with respect to the experimental results obtained in [1] using the expression (5) for imaginary part of the dielectric constant, which is responsible for losses. Rigorously speaking, the field absorption - losses $L$ - has to be calculated according to the expression:

$$L_{CNT} = 1 - \exp\left(2i\frac{\omega}{c}\operatorname{Im}\left[\sqrt{\varepsilon_1 + i\varepsilon_2}\,d\right]\right) \quad (6)$$

where $d$ is the effective thickness of the CNT layer. Positive values of $\varepsilon_2$ correspond to losses, negative – to amplification (the developed model allows us to consider both cases).

First, the effect of the enhancement of the nonlinearity can be now described quantitatively based on the elaborated expression (6).

Assuming that the imaginary part is smaller compare to the real one, we expand (6) into series:

$$L_{CNT} = 1 - \exp\left(2i\frac{\omega}{c}\operatorname{Im}\left[\sqrt{\varepsilon_1 + i\varepsilon_2}\,d\right]\right) \approx 1 - \exp\left(-\frac{\omega}{c}\frac{\varepsilon_2}{\sqrt{\varepsilon_1}}d\right) \quad (7)$$

In order to be close to the experimental procedure in [1], we considered relative change of absorption due to saturation in terms of the model parameters:

$$\frac{\Delta L_{CNT}}{L_{CNT}} \approx \frac{\omega d}{c}\left(\left.\frac{\varepsilon_2}{\sqrt{\varepsilon_1}}\right|_{|A|^2} - \left.\frac{\varepsilon_2}{\sqrt{\varepsilon_1}}\right|_0\right) \quad (8)$$

We can also assume (for simplicity) that the real part of the dielectric constant remains unchanged $\varepsilon_1|_{|A|^2} = \varepsilon_1|_0 = \varepsilon_{1,0}$:

$$\frac{\Delta L_{CNT}}{L_{CNT}} = \frac{\omega d}{c\sqrt{\varepsilon_{1,0}}}\left(\varepsilon_2(|A|^2) - \varepsilon_2(0)\right) < 0 \quad (9)$$

First, the effect will be evaluated for the when the resonance frequencies of CNT and metamaterial coincide, namely $\omega = \omega_0 = \omega_{21}$. The frequency dependent coefficients in this case become:

$$\begin{cases} |R_x|^2 = 4\omega^2\gamma^2 \\ |R_\rho|^2 = 1 \\ F_1 = 0 \\ F_2 = 0 \\ F_3 = 0 \\ S = S_0\left[1 + \dfrac{\sigma^2}{4\omega^2\gamma^2}\right] \end{cases} \quad (10)$$

and the relative absorption change (11) becomes (no pumping of CNT, $N_0 = -1$):

$$\left(\frac{\Delta L_{CNT}}{L_{CNT}}\right)_{Resonance} = -\frac{\omega d}{c\sqrt{\varepsilon_1}}\frac{4\pi\tau_2\mu^2}{\hbar}\left(1 - \frac{1}{1+S}\right) \quad (11)$$

In order to visualize the effect of the nonlinear response enhancement, it is necessary to introduce a relative transmission change according to the following expression:

$$\frac{\left(\dfrac{\Delta L_{CNT}}{L_{CNT}}\right)_{\sigma \neq 0, Resonance}}{\left(\dfrac{\Delta L_{CNT}}{L_{CNT}}\right)_{\sigma = 0, Resonance}} = \frac{1 + S_0\left(1 + \dfrac{\sigma^2}{4\omega^2\gamma^2}\right)}{(1 + S_0)\left(1 + \dfrac{\sigma^2}{4\omega^2\gamma^2}\right)} \quad (12)$$

which clearly demonstrates that the relative increase of transmission for CNT in the presence of the metamaterial ($\sigma \neq 0$) is bigger than for CNT alone ($\sigma = 0$). The respective dependence is presented in Fig. 1.

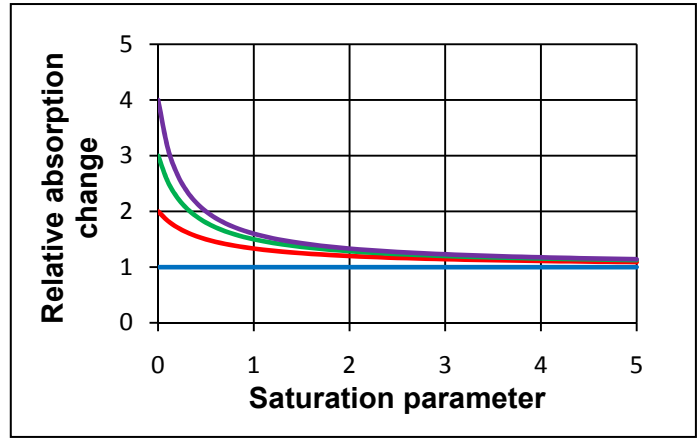

**Fig. 1: Relative transmission (12) as a function of saturation parameter** $S_0 = \dfrac{|A|^2}{|A_{s,0}|^2}$ **for different values of coupling** $\dfrac{\sigma^2}{4\omega^2\gamma^2} = 0, 1, 2, 3$ **("0" - blue line, "1" – red line, "2" – green line, "3" – violet line).**

From (12) we can conclude the following:

1. The coupling of CNT with the plasmonic nano-resonator enhances the CNT nonlinearity (saturation intensity becomes effectively lower).
2. Losses in the nano-resonator reduce the enhancement effect (higher $\gamma$ lead to higher saturation intensities).
3. The enhancement is more pronounced in the region of low saturation.

It is rather hard to estimate $\sigma$, but even under very conservative assumptions its value can easily become large then $\omega\gamma \sim 10^{28}$ and, consequently the effect of nonlinearity enhancement should be easily observed in the experiments.

It is important to discuss the influence of the Purcell effect [4] on the nonlinear properties of the considered hybrid structure. The Purcell effect appears when the density matrix elements, which are operators, become averaged over possible states of the system. These states are the combination of the states of CNT (assumed to be eigen states of CNT unperturbed Hamiltonian) and states of the electric field of the incident wave and that produced by the nano-resonator. In framework of our approach both fields contain huge number of photons and to this extend are safely classical, and the density matrix operator averaged over possible states becomes a function. The Purcell effect (increase of DOS of the field generated by the nano-resonator) is included in the constant $\alpha_x$, which is basically an integral over the mentioned possible states of the system, and consequently the only constant that can be changed. Given that the coupling constant $\sigma = \frac{\alpha_x \chi}{\mu_{CNT}}$ in (14) is proportional to $\alpha_x$ the Purcell effect cannot be distinguished from the field enhancement. There are two relaxation times in our model - $\tau_1, \tau_2$. Both appear as a result of interaction with the thermostat, i.e. with all possible fields except those produced by the nano-resonator (described by parameter $\alpha_x$), and therefore are not affected by Purcell effect.

**IV. Results of model**

In order to assess the validity of the proposed model the following strategy has been adopted. We compare the calculated and measured data for CNT layer alone and find CNT parameters from the fitting. Then we repeat the same procedure for the metamaterial without CNT to find parameters for the metamaterial. Next, we consider coupling between CNT and metamaterial structure and analyse the results for different values of the coupling constant $\sigma$. Direct comparison between calculated and measured data for the coupled system will be a subject of separated publication.

**CNT alone**

We first compared absorption spectrum of CNT layer itself with the experimental data. It turned out that the measured spectra exhibit visible asymmetry, which can be explained by the fact that the absorption lines of deposited CNT have different central frequencies. The effect of such inhomogeneous broadening is not included into the density matrix formalism and has to be taken into account by an additional averaging over respective probability distribution function (PDF). We assume in our modelling that PDF of CNT consists of two peaks, with both central frequency and bandwidth treated as the parameters that can be found from fitting the experimental spectrum.

The absorption was calculated according to the following expressions:

$$\begin{cases} L_{CNT,homogen}(\omega,\omega_{21}) = 1 - \exp\left(2i\frac{\omega}{c}\text{Im}\left[\sqrt{\varepsilon_1 + i\varepsilon_2}\right]d\right) \\ \varepsilon_{CNT} = \varepsilon_1 + i\varepsilon_2 \\ \varepsilon_1 = \varepsilon_{1,0} + \frac{4\pi n_{CNT}\tau_2^2\mu^2}{\hbar|R_\rho|^2(1+S)}(\omega - \omega_{21}) \\ \varepsilon_2 = \frac{4\pi n_{CNT}\tau_2\mu^2}{\hbar|R_\rho|^2(1+S)} \\ PDF(\omega_{21}) = C_1\exp\left(-\frac{(\omega_{21}-\omega_1)^2}{2\Delta\omega_1^2}\right) + C_2\exp\left(-\frac{(\omega_{21}-\omega_2)^2}{2\Delta\omega_2^2}\right) \\ L_{CNT,inhomogen}(\omega) = \int_0^\infty L_{I,CNT,homogen}(\omega,\omega_{21})*PDF(\omega_{21})d\omega_{21} \end{cases} \quad (15)$$

where the following values of the fitting parameters were used:

$\omega_1 = 9,9945\times 10^{14}$ (rad/s), $\omega_2 = 9,2174\times 10^{14}$ (rad/s), $\Delta\omega_1 = 1,65\times 10^{10}$ (rad/s), $\Delta\omega_2 = 5,2\times 10^{10}$ (rad/s)

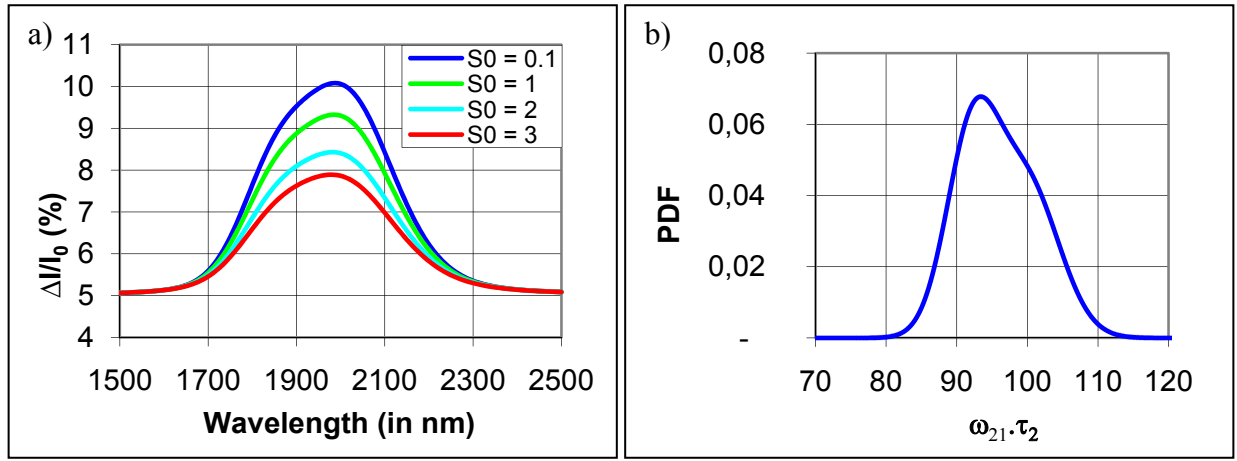

**Fig. 2: a) Calculated absorption spectrum of the CNT for different values of saturation parameter $S_0 = \frac{|A|^2}{|A_{s,0}|^2}$. The experimental data coincide one to one with the unsaturated line (blue line) and is not shown separately; b) The PDF used to calculate the absorption spectrum.**

**Metamaterial alone**

After the parameters of the CNT have been retrieved, one can obtain the modelling parameters of the metamaterial structure itself. The system of equations in this case does not include any extra averaging:

$$\begin{cases} L_{MM}(\omega, \omega_{21}) = 1 - \exp\left(2i\frac{\omega}{c}\operatorname{Im}\left[\sqrt{\varepsilon_1 + i\varepsilon_2}\right]d\right) \\ \varepsilon_x = \varepsilon_1 + i\varepsilon_2 \\ \varepsilon_1 = \varepsilon_{1,0} + \dfrac{4\pi n_{MM} q\chi(\omega_0^2 - \omega^2)}{|R_x|^2} \\ \varepsilon_2 = \dfrac{8\pi q n_{MM} \chi \omega \gamma}{|R_x|^2} \end{cases} \qquad (16)$$

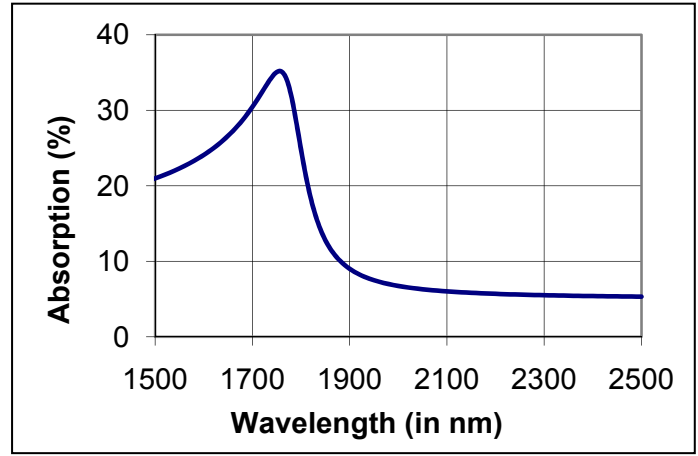

Fig. 3: Calculated absorption spectrum of the metamaterial.

Note that the calculated spectrum of the metamaterial does not match the experimental one. This is because we used a very simple model for the plasmonic metamaterial based on a single harmonic oscillator and aimed only to reproduced 30% level of absorption measured experimentally at the metamaterial resonance. As it can be seen from (19) the spectrum dependence of the absorption is stipulated by the function $|R_x|^2$, and in order to achieve a good match between the observed and calculated metamaterial spectral response it is necessary to either create an appropriate model based on the set of several coupled oscillators [5], or just choose an appropriate fitting function for $|R_x|^2$. Here we limit ourselves to demonstrating the adequateness of our very simple model. Correspondingly, in what follows we exploit the calculated absorption spectrum of the metamaterial structure shown in Fig. 3. It is worth noting that in the presence of CNT layer, the central frequency of the metamaterial absorption peak will shift reflecting the fact that the plasmon resonance frequency depends on the dielectric constant of environment. This effect is not included in our model, where central frequency $\omega_0$ is a fitting parameter.

**CNTs combined with metamaterial**

Here we present the results of modelling the enhancement of CNT nonlinearity response based on the values of fitting parameters obtained in two previous sections. The only parameter that could not be retrieved using the experimental data is the coupling constant $\sigma$, calculating which is the subject of a separate publication.

We consider two possible realisations of CNT coating, namely: with purely homogeneous distribution of CNT parameters over the metamaterial surface (PDF in Fig. 2-b becomes delta function), and inhomogeneous, when each CNT has different eigen frequency and oscillates with eigen phase shifted relative to the incoming field and local field of the metamaterial nano-resonator. In homogeneous case we do not take into account spatial distribution of the CNT eigen frequency in the calculation of absorption, which means that we can simply add effective dielectric constants of CNT and metamaterial layers. In inhomogeneous case the contribution to the absorption from the CNT molecules becomes not fully coherent and has to be taken into account separately from the metamaterial one. This can be done by summation of the refractive indexes of CNT and metamaterial instead of summation of their dielectric constants, as in coherent case.

As can be seen from Fig. 1, the effect of the enhancement is mostly pronounced for small saturation values, and may be observed in almost complete absence of the metamaterial if saturation parameter exceeds 3.

The homogeneous case is described by the following system of equations:

$$\begin{cases} L_{CNT\&MM,homo}(\omega,\omega_{21}) = 1 - \exp\left(2i\frac{\omega}{c}\text{Im}\left[\sqrt{\varepsilon_1 + i\varepsilon_2}\right]d\right) \\ \varepsilon_{CNT} + \varepsilon_x = \varepsilon_1 + i\varepsilon_2 \\ \varepsilon_1 = \varepsilon_{1,0} - \frac{4\pi n_{CNT}\tau_2\mu^2}{\hbar|R_\rho|^2(1+S)}\left((\omega-\omega_{21})\tau_2 + \frac{2\omega\gamma\sigma}{|R_x|^2}\right) + \frac{4\pi n_{MM}q\chi(\omega_0^2 - \omega^2)}{|R_x|^2} \\ \varepsilon_2 = \frac{4\pi n_{CNT}\tau_2\mu^2}{\hbar|R_\rho|^2(1+S)}\left(1 + \frac{\sigma F_1}{|R_x|^2}\right) + \frac{8\pi q n_{MM}\chi\omega\gamma}{|R_x|^2} \end{cases} \quad (17)$$

while for incoherent case:

$$\begin{cases} L_{CNT\&MM,incoh}(\omega,\omega_{21}) = \\ = 1 - \exp\left(2i\frac{\omega}{c}\operatorname{Im}\left[\sqrt{\varepsilon_{1,CNT\_A} + i\varepsilon_{2,CNT\_A}} + \sqrt{\varepsilon_{1,CNT\_x} + i\varepsilon_{2,CNT\_x}} + \sqrt{\varepsilon_{1,x} + i\varepsilon_{2,x}}\right]d\right) \\ \varepsilon_{1,CNT\_A} = \varepsilon_{1,0} - \frac{4\pi n_{CNT}\tau_2\mu^2}{\hbar|R_\rho|^2(1+S)}(\omega - \omega_{21})\tau_2 \\ \varepsilon_{2,CNT\_A} = \frac{4\pi n_{CNT}\tau_2\mu^2}{\hbar|R_\rho|^2(1+S)} \\ \varepsilon_{1,CNT\_x} = -\frac{4\pi n_{CNT}\tau_2\mu^2}{\hbar|R_\rho|^2(1+S)} \times \frac{2\omega\gamma\sigma}{|R_x|^2} \\ \varepsilon_{2,CNT\_x} = \frac{4\pi n_{CNT}\tau_2\mu^2}{\hbar|R_\rho|^2(1+S)} \times \frac{\sigma F_1}{|R_x|^2} \\ \varepsilon_{1,x} = \frac{4\pi n_{MM}q\chi(\omega_0^2 - \omega^2)}{|R_x|^2} \\ \varepsilon_{2,x} = \frac{8\pi q n_{MM}\chi\omega\gamma}{|R_x|^2} \end{cases} \quad (18)$$

Here $\varepsilon_{1,CNT\_A}, \varepsilon_{2,CNT\_A}$ comes from CNT driven by the external field, $\varepsilon_{1,CNT\_x}, \varepsilon_{2,CNT\_x}$ is the contribution due to the local field of the metamaterial's nano-resonators, and $\varepsilon_{1,x}, \varepsilon_{2,x}$ are the effective dielectric constants of the metamaterial.

Fig. 4 presents the results of our calculations for the homogeneous and inhomogeneous cases with the saturation parameter set to 3.

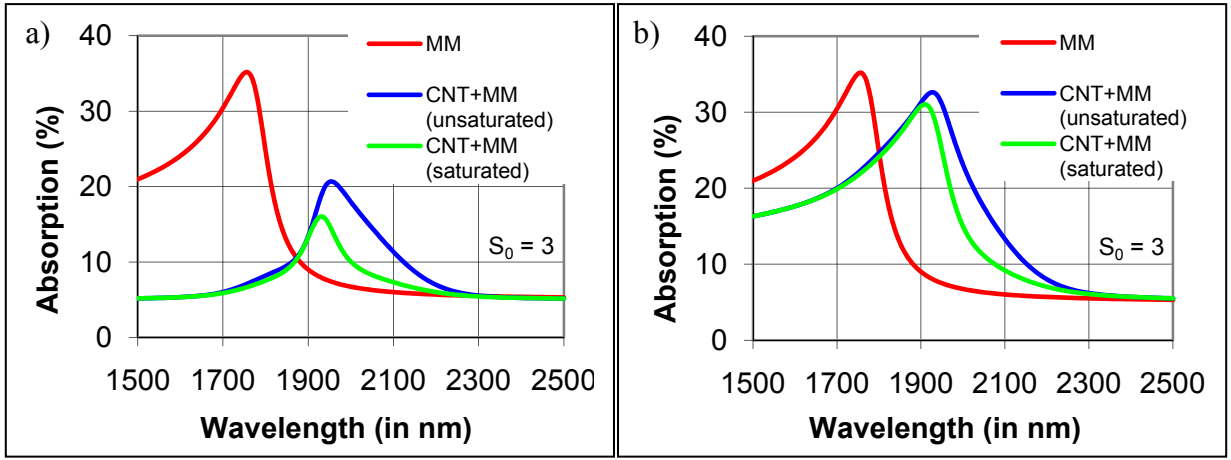

**Fig. 4: Absorption spectrum of CNTs combined with metamaterial for a) homogeneous and b) inhomogeneous cases.**
Saturation parameter $S_0 = \frac{|A|^2}{|A_{s,0}|^2} = 3$.

For comparison we also present here the absorption of the metamaterial alone (red curve in Fig. 4). The absorption spectra of the CNT-coated metamaterial are red shifted with respect to the uncoated structure due to the fact that CNT effectively increase the density of the dielectric environment for the nano-resonators. In order to demonstrate the effect of the metamaterial on the absorption change due to saturation, we plotted the normalised absorption change $\frac{\Delta L_{CNT\&MM}}{L_{CNT\&MM}}$, shown in Fig. 5.

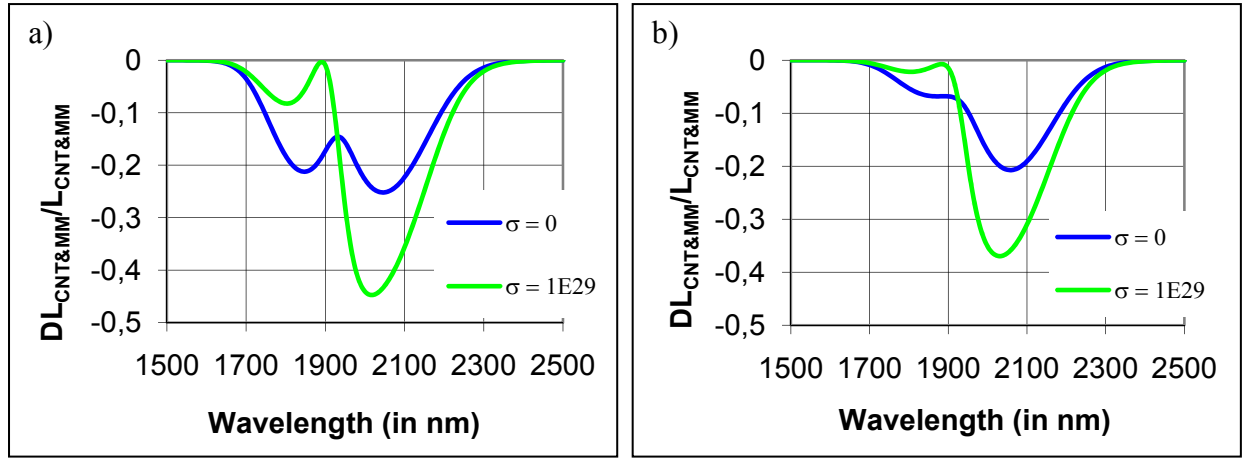

Fig. 5: Normalised absorption change spectrum of CNTs combined with metamaterial for a) homogeneous and b) inhomogeneous cases. Saturation parameter is $S_0 = \frac{|A|^2}{|A_{s,0}|^2} = 3$.

The Fig. 5 clearly demonstrates that in homogeneous case the effect appears to be more pronounced, which leads to the general requirement of maximizing the homogeneity of the deposited layer of CNT.

**Conclusion**

A simple and robust model describing the enhancement of saturation nonlinearity in the system of coupled CNT and metamaterial is developed. The model adequately reproduces absorption spectrum of the CNT itself taking into account distribution of center frequency of the CNT absorption line. The model clearly demonstrates the effect of saturation nonlinearity enhancement due to the presence of the metamaterial, and the dependence of the effect on homogeneity of the CNT layer.